\documentclass[preprint,showkeys,secnumarabic,showpacs,amsmath,amssymb,superscriptaddress,]{revtex4}
\usepackage{graphicx}
\begin{document}
\title{Warm constant-roll inflation in brane-world cosmology}

\author{M. R. Setare}
\email{rezakord@ipm.ir}
\affiliation{Department of Science, Campus of Bijar, University of Kurdistan, Bijar, Iran}

\author{A. Ravanpak}
\email{a.ravanpak@vru.ac.ir}
\affiliation{Department of Physics, Vali-e-Asr University of Rafsanjan, Rafsanjan, Iran}

\author{K. Bahari}
\email{karam.bahari@gmail.com}
\affiliation{Physics Department, Faculty of Science, Razi University, Kermanshah, Iran}

\author{G. F. Fadakar}
\email{g.farpour@vru.ac.ir}
\affiliation{Department of Physics, Vali-e-Asr University of Rafsanjan, Rafsanjan, Iran}

\date{\small {\today}}

\begin{abstract}

In this article we study a constant-roll inflationary model in the context of brane-world cosmology caused by a quintessence scalar field for warm inflation with a constant dissipative parameter $Q =\Gamma/3H$. We determine the analytical solution for the Friedman equation coupled to the equation of motion of the scaler field. The evolution of the primordial scalar and tensor perturbations is also studied using the modified Langevin equation. To check the viability of the model we use numerical approaches and plot some figures. Our results for the scalar spectral index and the tensor to scaler ratio show good consistency with observations.

\end{abstract}

\pacs{98.80.Cq; 11.25.-w}

\keywords{constant-roll inflation; warm inflation; brane-world, perturbation.}

\maketitle

%----------------------------------------------------------------------------------------------------------------------------------------------------------------------------------
\section{Introduction}

Recent theoretical developments and precise astronomical data lead to considering an inflationary phase at some stage in the very early Universe that is one of the most convincing solution to many long-standing problems of the standard hot big-bang model such as the flatness, the horizon, and the monopole problems \cite{Guth}-\cite{Albrecht}. Besides, this inflationary era provides a causal interpretation of the origin of the measured anisotropy of the cosmic microwave background (CMB) radiation, and also the distribution of large scale structure of the Universe \cite{Liddle}-\cite{Hinshaw}.

Generally, inflation drives by the potential of an isolated standard scalar field called inflaton, though there are some other non-standard scalar field inflationary models \cite{Picon}. Various kinds of scalar fields can play the role of the inflaton field. For example the quintessence field and the tachyon field, among others. To support a long-enough period of inflation, they must roll down their potential very slowly. The process which is named slow-roll inflation. There are lots of articles about the slow-roll inflationary paradigm in the literature \cite{Faraoni}-\cite{Ravanpak}.

Although the slow-roll inflationary model which entails an approximately flat inflaton potential yields viable observational predictions, but many attempts have been directed towards exploring an inflationary solution in which the assumption of the inflaton slow-roll, has been omitted \cite{Starobinsky}-\cite{Namjoo}. Among them, the case studied in \cite{Tsamis}-\cite{Namjoo}, which is dubbed ultra-slow-roll inflation, is the most important one, because after its generalization in \cite{Martin}, the idea of constant-roll inflation was raised and developed \cite{Motohashi}-\cite{Oikonomou2}. Maybe the most interesting property of the constant-roll inflationary model is the prediction of non-Gaussianities \cite{Namjoo}-\cite{Motohashi2},\cite{Cai}.

All of the above inflationary scenarios, irrespective of slow-roll or constant-roll, are called cold inflation, since in all of them the isolated inflaton does not interact with other particle fields during inflation and therefore the temperature of the Universe decreases rapidly. Thus to connect the end of inflation with the radiation dominated big-bang phase we need a process called reheating. Usually, the inflaton field oscillates around the minimum of its potential and reheats the Universe. But, in the situations that the inflaton potential does not have a minimum, other mechanisms such as the curvaton scenario have to be used to bring inflation to an end \cite{Mollerach}-\cite{Campo}.

Warm inflation is another dynamical realization of inflation in which the inflaton field is not isolated anymore. In this scenario, the interaction between the inflaton field and other fields implies an effective evolution equation that contains a friction term which describes the production of radiation during inflationary expansion \cite{Berera}-\cite{Moss}. So, the reheating process is no longer needed. The friction term contains a dissipation coefficient that characterizes two high and weak dissipative regimes.

Independent of the above subjects, recently, higher dimensional theories of gravity, in which our Universe is considered as a brane embedded in a higher dimensional space-time, have attracted a great deal of attention because they may provide the solution for long-standing problems in cosmology. Based on these brane-world models, the standard model of particle physics is confined to a four-dimensional brane, whereas gravity can leak into the extra dimension. Thus they can explain the weakness of gravity and hierarchy problem \cite{Arkani}-\cite{Randall2}. The effect of the extra dimension gives rise additional terms in the Friedmann equation on the brane \cite{Binetruy}-\cite{Shiromizu}. Specially, the presence of a quadratic term of the energy density, makes it easier to explain inflation in the very early Universe \cite{Mohapatra},\cite{Maartens}.

Although many articles in the literature have studied various kinds of inflationary scenarios mentioned above separately \cite{Herrera}-\cite{Ghersi}, but there can be found some mixed situations. For example in \cite{Cid}, the authors have studied a warm slow-roll inflationary model in a high-dissipation regime on a brane scenario. They have shown the role of the dissipation parameter in producing the isocurvature and adiabatic modes of perturbations. Also, in \cite{Kamali}, the authors have investigated a warm constant-roll inflationary paradigm in a high dissipative regime using a new approach. Considering a constant dissipative parameter, they have studied the evolution of the primordial inhomogeneities of a scalar field in a thermal bath and then they have shown the consistency between the theoretical predictions of their model and observations.

In this article we will study a more complex mixed model in which the idea of a warm constant-roll inflation will be investigated in the context of the brane-world cosmology. Our main motivation to choose a brane context, regardless of the advantages and properties mentioned above is the latest results of Planck satellite. Because in \cite{Akrami}, the authors discriminate some of the inflationary models that are best able to account for the data and one of them is the so called D-brane inflation. We consider a quintessence scalar field as the inflaton field and also assume a constant dissipative parameter. We obtain the evolutionary equations of the model and solve them analytically. The evolution of the primordial inhomogeneities will be also studied. We utilize some plots to demonstrate the viability of our model. The structure of the article is as following: in Sec.2, we introduce the model, obtain the equations and solve them, analytically. Sec.3, deals with the perturbation theory. In this section we calculate the most important perturbation parameters of the model. Numerical discussions presents in Sec.4 in which we try to show the consistency between our results and observations. Finally, a summary and the conclusions are presented in Sec.5.

%-----------------------------------------------------------------------------------------------------------------------------------------------------------------------------------
\section{warm constant-roll inflation on the brane}

We consider a 5D brane cosmology in which the modified Friedmann equation on the brane is given by:
\begin{equation}\label{Friedmann}
    3H^2=\rho\left(1+\frac{\rho}{2\lambda}\right)+\Lambda+\frac{3\Xi}{a^4}
\end{equation}
Here, $a$, $H=\dot a/a$ (the dot means derivative with respect to the cosmic time $t$), $\rho$ and $\Lambda$, are the scale factor, the Hubble parameter, the energy density of matter content of the Universe and the cosmological constant, respectively, and we set the reduced Planck mass $m_p=1/\sqrt{8\pi G}=1$. The last term represents the influence of the bulk gravitons on the brane with $\Xi$ as an integration constant, called dark radiation because it scales as $a^{-4}$. Since inflation is a period in the very early universe, we impose the high energy condition $\rho\gg\lambda$, in our model. Also we assume that $\Lambda=0$ and just when the inflation begins the last term in Eq.(\ref{Friedmann}) will rapidly become unimportant. All these circumstances leave us with the following Friedmann equation:
\begin{equation}\label{FriedmannHE}
    3H^2=\frac{\rho^2}{2\lambda}
\end{equation}
Although in a cold inflationary scenario one can assume that the Universe is only filled by a homogeneous scalar field $\phi$, with the energy density $\rho_\phi=\dot\phi^2/2+V(\phi)$ ($V$ stands for potential), but in a warm inflationary case we need to consider the contribution of the radiation field energy density $\rho_r$, as well. So, we obtain the following effective Friedmann equation:
\begin{equation}\label{FriedmannHE2}
    3H^2=\frac{(\rho_\phi+\rho_r)^2}{2\lambda}
\end{equation}
and also
\begin{equation}\label{Raychaudhuri}
    \dot H=-\sqrt{\frac{3}{2\lambda}}H\left(\dot\phi^2+\frac{4}{3}\rho_r\right)
\end{equation}
Meanwhile, the dissipation between energy densities of the inflaton field and radiation, modifies continuity equations as
\begin{eqnarray}
\ddot\phi+(3H+\Gamma)\dot\phi+V_\phi &=& \ddot\phi+3H(1+Q)\dot\phi+V_\phi=0 \\
\dot\rho_r+4H\rho_r&=&\Gamma\dot\phi^2=3HQ\dot\phi^2 \label{rhor}
\end{eqnarray}
in which $\Gamma=3HQ$ is the dissipation coefficient and $V_\phi$, represents derivative with respect to $\phi$.

The condition of constant-roll inflation can be stated as below
\begin{equation}\label{cr}
\ddot\phi+\beta H\dot\phi=0
\end{equation}
which yields
\begin{equation}\label{dphi}
\dot\phi=\dot\phi_0e^{-\beta N}
\end{equation}
where we have used the relation $a=e^N$, in which $N$ is the number of $e$-folds. It is obvious that for $\beta=0$, the standard slow-roll inflation occurs.

Using Eq.(\ref{dphi}), we can solve Eq.(\ref{rhor}) analytically for $Q=constant$, as
\begin{equation}\label{rho}
\rho_r=\rho_{r0}a^{-4}+\frac{3Q\dot\phi_0^2}{2(2-\beta)}a^{-2\beta}
\end{equation}
One can employ Eq.(\ref{rho}) to modify Eq.(\ref{Raychaudhuri}), which takes the form of
\begin{equation}\label{H}
\frac{\sqrt{6\lambda}}{2}\frac{dH^2}{dN}=-\left(3H\dot\phi_0^2 e^{-2\beta N}+4H\rho_{r0}e^{-4N}+\frac{6QH\dot\phi_0^2}{2-\beta}e^{-2\beta N}\right)
\end{equation}
By integrating both sides of Eq.(\ref{H}) with respect to $N$, one obtains the following equation:
\begin{equation}\label{H1}
\sqrt{6\lambda}H=\frac{3\dot\phi_0^2e^{-2\beta N}}{2\beta}\left(\frac{2+2Q-\beta}{2-\beta}\right)+\rho_{r0}e^{-4N}+V_0
\end{equation}
in which $V_0$, is a constant of integration. One can interpret $V_0$, as a constant part of the inflationary potential or as a cosmological constant. Comparing with the Friedmann equation (\ref{FriedmannHE2}), we can extract the potential $V$ as below:
\begin{equation}\label{V}
V(N)=\frac{\dot\phi_0^2e^{-2\beta N}}{2\beta}(3Q+3-\beta)+V_0
\end{equation}
In order to have a positive potential for all values of $N$, we need $\beta\leq3(1+Q)$. This is consistent with $dV/dN\leq0$, which is the condition requires for the inflaton field.

In order to obtain the potential as a function of the inflaton field, we need to find $N=N(\phi)$. Considering
\begin{equation}\label{N}
\dot\phi=H\frac{d\phi}{dN}=\dot\phi_0e^{-\beta N}
\end{equation}
and using Eq.(\ref{H1}) we obtain
\begin{equation}\label{phiN}
\phi(N)={\sqrt{6\lambda}\dot\phi_0}\int{\frac{e^{-\beta N}dN}{\frac{3\dot\phi_0^2e^{-2\beta N}}{2\beta}(\frac{2+2Q-\beta}{2-\beta})+\rho_{r0}e^{-4N}+V_0}}
\end{equation}
For general values of $\beta$, the solution for $\phi(N)$ cannot be found. But, if we ignore $\rho_{r0}e^{-4N}$ and $V_0$, compared to the first term in the denominator, we find
\begin{equation}\label{p}
\phi-\phi_0=\frac{2\sqrt{2\lambda}}{\sqrt3\dot\phi_0}\left(\frac{2-\beta}{2+2Q-\beta}\right)e^{\beta N}
\end{equation}
so that
\begin{equation}\label{N}
e^{-2\beta N}=\left(\frac{2\sqrt{2\lambda}}{\sqrt3\dot\phi_0}\left(\frac{2-\beta}{2+2Q-\beta}\right)\right)^2(\phi-\phi_0)^{-2}
\end{equation}
Then the scalar field potential equals
\begin{equation}\label{Vphi}
V(\phi)=\frac{4\lambda(2-\beta)^2(3Q+3-\beta)}{3\beta(2+2Q-\beta)^2}(\phi-\phi_0)^{-2}
\end{equation}
Similarly, we can solve the Eq.(\ref{phiN}) analytically, by just assuming that, $\rho_{r0}=0$, which means that there is
no radiation other than the one produced by dissipation. In this case we obtain
\begin{equation}\label{v1}
\phi-\phi_0=2\sqrt{\frac{\lambda}{\beta V_0}\left(\frac{2-\beta}{2+2Q-\beta}\right)}\arctan\left(\frac{e^{\beta N}}{\dot\phi_0}\sqrt{\frac{2V_0\beta}{3}\left(\frac{2-\beta}{2+2Q-\beta}\right)}\right)
\end{equation}
which yields
\begin{equation}\label{v}
e^{-2\beta N}=\left(\frac{2V_0\beta}{3\dot\phi_0^2}\left(\frac{2-\beta}{2+2Q-\beta}\right)\right)\left(\tan\left(\left(\frac{\phi-\phi_0}{2}\right)\sqrt{\left(\frac{\beta V_0}{\lambda}\right)\left(\frac{2+2Q-\beta}{2-\beta}\right)}\right)\right)^{-2}
\end{equation}
Using the equation above  one can find the potential
\begin{equation}\label{v2}
V(\phi)=\left(\frac{{V_0(2-\beta)(3Q+3-\beta)} }{3{(2+2Q-\beta)}}\right)\left(\tan\left(\left(\frac{\phi-\phi_0}{2}\right)\sqrt{\left(\frac{\beta V_0}{\lambda}\right)\left(\frac{2+2Q-\beta}{2-\beta}\right)}\right)\right)^{-2}+V_0
\end{equation}

%We know that in the usual inflationary scenario, the inflaton field decays to radiation at the end of inflation and reheats the universe, but in the warm inflationary scenario, it decays during inflation.

To obtain the Hubble parameter as a function of the inflaton field we substitute $N$ from Eq.(\ref{V}) into Eq.(\ref{H1}). Since we have neglected the term containing $\rho_{r0}$, for both cases of $V_0=0$ and $V_0\neq 0$ we arrive at
\begin{equation}\label{Hphi}
H=\frac{3}{\sqrt{6\lambda}}\frac{(2-\beta+2Q)(V(\phi)-V_0)}{(2-\beta)(3-\beta+3Q)}.
\end{equation}

%----------------------------------------------------------------------------------------------------------------------------------------------------
\section{warm inflation perturbations}

In this section we study the perturbations in warm inflation and try to investigate these perturbations analytically. Here we follow the method used in \cite{Kamali}. Therein, the authors mentioned that the evolution of the inflaton field is governed by the modified Langevin equation:
\begin{equation}\label{Eq Langevian}
-\square\phi(x,t)+\Gamma\dot{\phi}+V_\phi=(2\Gamma_{eff}T)^{1/2}\xi(x,t)
\end{equation}
Here, $\square$ is the Laplace operator in the four dimensional space-time, $\Gamma_{eff}=H(1+Q)$ and $\xi$ is a stochastic variable with approximately
Gaussian distribution. The two-point correlation function is given by:
\begin{equation}\label{Eq correlation 1}
\langle\xi(x,t),\xi(x',t')\rangle=\delta^3(x-x')\delta(t-t')
\end{equation}
The inflaton field can be considered as the superposition of a homogeneous term and a perturbation term as
\begin{equation}\label{superposition}
\phi(x,t)=\phi(t)+\delta\phi(x,t)
\end{equation}
and the Gaussian condition can be rewritten as
\begin{equation}\label{Eq correlation 2}
\langle\xi(x,t),\xi(x',t')\rangle=a^{-3}(2\pi)^2\delta^3(x-x')\delta(t-t').
\end{equation}

Assuming $T>H$ \cite{Graham}, the perturbation part of the Langevin equation after Fourier transformation becomes
\begin{equation}\label{Eq Langevian Furier}
\delta\ddot{\phi}(\mathbf{k},t)+(3H+\Gamma)\delta\dot{\phi}(\mathbf{k},t)+V_{\phi\phi}\delta\phi(\mathbf{k},t)+k^2a^{-2}\delta\phi(\mathbf{k},t)=
(2\Gamma_{eff}T)^{1/2}\xi(\mathbf{k},t)
\end{equation}

If we use the new variable $z=k/aH$ instead of time variable the above equation transforms to
\begin{equation}\label{Eq Perturbed Langevian}
\Big(1+\frac{\dot{H}}{H^2}\Big)^2\delta\phi''-\Big(\frac{3Q+2}{z}\Big)\Big(1+\frac{\dot{H}}{H^2}\Big)\delta\phi'+
\Big(1+\frac{\dot{H}}{H^2}\Big)\Big(\frac{\dot{H}}{H^2}\Big)'\delta\phi'+\frac{V_{\phi\phi}}{z^2H^2}\delta\phi+\delta\phi=\\
(2\Gamma_{eff}T)^{1/2}\Big(\frac{a}{k}\Big)^2\xi(\mathbf{k},z)
\end{equation}
in which the prime represents differentiation with respect to $z$. Unfortunately this equation is untractable analytically for nonzero $V_0$, hence hereafter we focus on the case $V_0=0$ for which an analytical solution can be determined. For this case we find that $\varepsilon=-\dot{H}/H^2=2\beta$ which in turn results an inflationary solution only for $\beta\ll1$. So, our model for $V_0=0$ is close to a standard slow-roll inflationary scenario. In such a case, one can check that $\frac{V_{\phi\phi}}{H^2}=3\beta(3Q+3-\beta)$. Using these relations Eq.(\ref{Eq Perturbed Langevian}) can be rewritten as
\begin{equation}\label{Eq Perturbed Langevian 1}
(1-2\beta)^2\delta\phi''-\frac{(3Q+2)(1-2\beta)}{z}\delta\phi'+
\frac{3\beta(3Q+3-\beta)}{z^2}\delta\phi+\delta\phi=
(2\Gamma_{eff}T)^{1/2}\Big(\frac{a}{k}\Big)^2\xi(\mathbf{k},z)
\end{equation}
If we substitute $z\rightarrow z(1-2\beta)$, more simplifications can be achieved and we arrive at
\begin{equation}\label{Eq Perturbed Langevian 2}
\delta\phi''-\Big[\frac{3Q+2}{z(1-2\beta)}\Big]\delta\phi'+
\frac{3\beta(3Q+3-\beta)}{(1-2\beta)^2z^2}\delta\phi+\delta\phi=
(2\Gamma_{eff}T)^{1/2}\Big(\frac{a}{k}\Big)^2\xi(\mathbf{k},z).
\end{equation}
The homogeneous counterpart of Eq.(\ref{Eq Perturbed Langevian 1}) is the same as Eq.(\ref{Eq generalized Bessel equation}) in the appendix with $\gamma=1$, $\theta=1$, $\nu=\frac{1}{2}\big(1+\frac{3Q+2}{1-2\beta}\big)\simeq\frac{3}{2}(Q+1)$ and
\begin{equation}\label{Eq n}
n^2=\nu^2-\frac{3\beta(3Q+3-\beta)}{(1-2\beta)^2}\simeq\nu^2-9\beta(1+Q)
\end{equation}
The expression for $n$ can be more simplified as $n=\pm(\nu+3c)$ in which $c=-\beta$. It can be shown that the negative $n$, yields a blue-tilted spectrum of scalar fluctuations that is ruled out by observations \cite{Kamali}. In the following we study the case $n=\nu+3c$. The solution of Eq.(\ref{Eq Perturbed Langevian 2}) is given in terms of the Green function as
\begin{equation}\label{Eq Solution Using Green Function}
\delta\phi=\int_z^{\infty}G(z,z')(z')^{1-2\nu}{\xi}(z')(2\Gamma_{eff}T)^{1/2}dz',
\end{equation}
in which the Green function is given by
\begin{equation}\label{Eq Green Function}
G(z,z')=\frac{\pi}{2}z^\nu {z'}^{\nu}(J_{\nu+3c}(z)Y_{\nu+3c}(z')-J_{\nu+3c}(z')Y_{\nu+3c}(z))
\end{equation}
The power spectrum of the inflaton field is related to the two-point correlation function via
\begin{equation}\label{Eq correlation 2}
\langle\delta\phi(k,z)\delta\phi(k',z)\rangle=P_{\phi}(2\pi k)^3\delta^3(k+k')
\end{equation}
So, using Eqs.(\ref{Eq Solution Using Green Function}), (\ref{Eq Green Function}) and (\ref{Eq correlation 2}), we arrive at:
\begin{equation}\label{Eq power spectrum 1}
P_{\phi}=(2\Gamma_{eff}T)\int_z^{\infty}{(z')}^{2-4\nu}G(z,z')^2dz'
\end{equation}
For the values of $z$ much smaller than unity the power spectrum can be simplified to
\begin{equation}\label{Eq power spectrum 2}
P_{\phi}=(2\Gamma_{eff}T)\frac{\Gamma_R(c+3/2)}{\Gamma_R(3/2)}\sqrt{\frac{\pi}{32\nu}}\Big(\frac{2\nu}{z^2}\Big)^{3c}.
\end{equation}
Here $\Gamma_R$ is the Gamma function. The curvature perturbation is defined as \cite{Graham}
\begin{equation}\label{Eq Curvature perturbation 1}
\zeta=\frac{1}{2}\ln(1+2\varphi)+\frac{1}{3}\int\frac{d\rho}{\rho+p}
\end{equation}
in which $p$ and $\varphi$ represents the pressure and the spatial curvature perturbation, respectively. In high dissipative warm constant-roll situation, it can be modified to
\begin{equation}\label{Eq Curvature perturbation 2}
\zeta=\frac{1}{2}\ln(1+2\varphi)-\int\frac{H}{\dot{\phi}}d\phi
\end{equation}
Assuming a uniform curvature gauge $\varphi=0$ and $\zeta\equiv\zeta(\phi)$, and with attention to Eq.(\ref{superposition}) one obtains
\begin{equation}\label{Eq Curvature perturbation 3}
\zeta=\zeta_\phi\delta\phi
\end{equation}
in which $\zeta_\phi=-\frac{H}{\dot{\phi}}=-\frac{1}{\beta}\phi^{-1}$. So, the power spectrum of density perturbations is
\begin{equation}\label{Eq P xi}
P_{\zeta}=\zeta_\phi^2P_{\phi}=\Big(\frac{H}{\dot{\phi}}\Big)^2P_{\phi}.
\end{equation}
The spectral index is then given by
\begin{equation}\label{Eq spectral index}
n_s-1=\frac{d\ln P_{\zeta}}{d\ln k}=-2\varepsilon=-4\beta.
\end{equation}

In addition to scalar perturbations, the concept of tensor perturbations is important. Specifically, in a brane-world scenario tensor perturbations are more involved because gravitons can propagate in the bulk. We can express the amplitude of the tensor perturbations as below \cite{Langlois}:
\begin{equation}\label{Tperturbation}
P_g=8\left(\frac{H}{2\pi}\right)^2F^2(x)
\end{equation}
in which $x=H\sqrt{3/(4\pi\lambda)}$, and
\begin{equation}\label{Fx}
F(x)=\left[\sqrt{1+x^2}-x^2\sinh^{-1}(\frac{1}{x})\right]^{-\frac{1}{2}}
\end{equation}
We have to note that the thermal bath of warm inflation does not affect the tensor perturbations \cite{Taylor}. Finally, using equations above we obtain the tensor to scaler ratio as
\begin{equation}\label{Eq t-to-s ratio}
r=\frac{P_g}{P_{\zeta}}=\frac{1}{\pi^2}\frac{\Gamma_R(\frac{3}{2})}{\Gamma_R(c+\frac{3}{2})}\frac{(32\nu)^{\frac{1}{2}}}{(2\nu)^{3c}\sqrt{3\pi}}\frac{(2-\beta)}{(2+2Q-\beta)}\frac{2\beta\sqrt{2\lambda}}{(1+Q)T}F^2(x)
\end{equation}

%--------------------------------------------------------------------------------------------------------------------------------------------------------------------------------
\section{numerical discussions}

We can constrain our model parameters comparing the results calculated here with those obtained observationally by Planck 2018 data \cite{Akrami},\cite{Aghanim}. The Planck collaboration has been realized that in a $\Lambda$CDM+$r$ model, the scalar spectral index takes the value $n_s=0.9668\pm0.0037$ in the 68\% confidence limit. So, with attnetion to Eq.(\ref{Eq spectral index}) we can constrain the constant-roll parameter $\beta$, as $0.0074<\beta<0.0092$.
%Also, they have evaluated an upper bound on the tensor to scalar ratio at the scale $k=0.002$ Mpc$^{-1}$ in the 95\% confidence limit as $r_{0.002}<0.058$.

\begin{figure}[h]
\centering
\includegraphics[width=8cm]{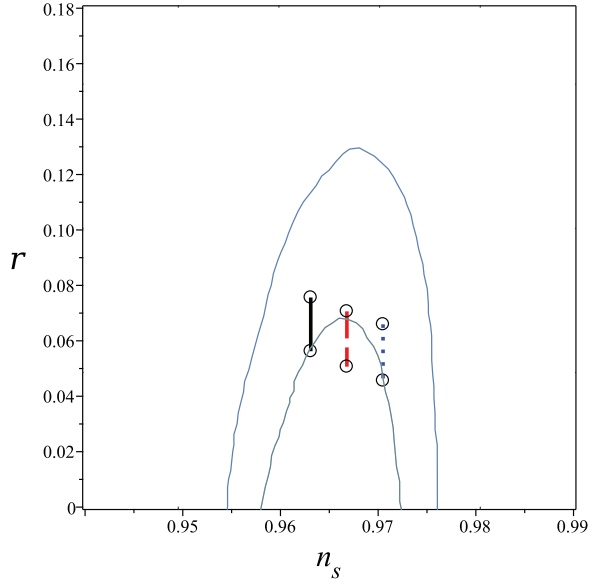}
\caption{The trajectories in $r-n_s$ plane. The black solid line, the red dashed line and the blue dotted line have been plotted for $\beta=0.0074$, $\beta=0.0083$ and $\beta=0.0092$, respectively. Also, in each case the upper circle relates to $N=50$ while the lower one relates to $N=70$.}\label{fig1}
\end{figure}

\begin{figure}[h]
\centering
\includegraphics[width=8cm]{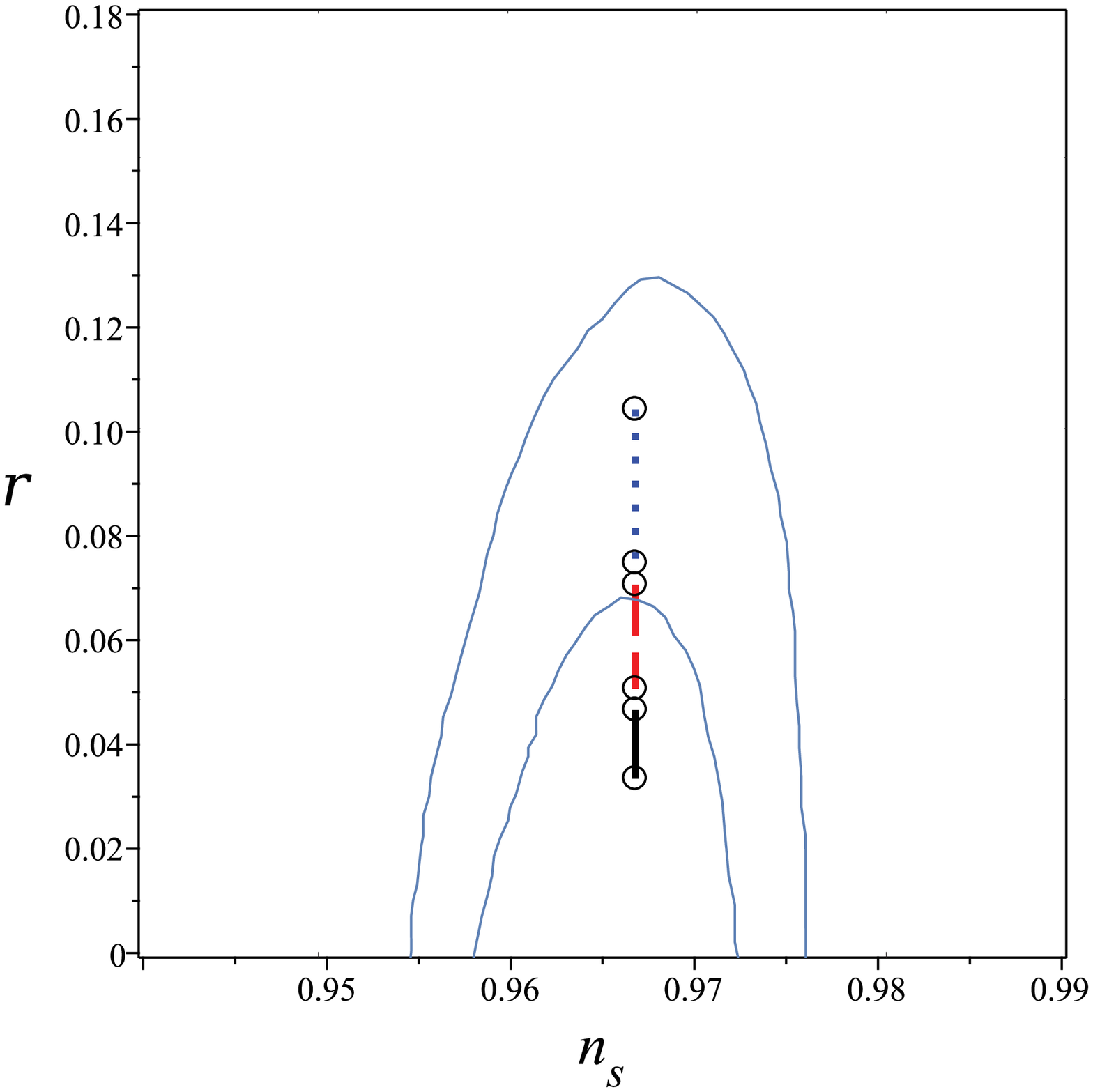}
\caption{The trajectories in $r-n_s$ plane. The black solid line, the red dashed line and the blue dotted line have been plotted for $Q=2400$, $Q=1000$ and $Q=440$, respectively. Also, in each case the upper circle relates to $N=50$ while the lower one relates to $N=70$.}\label{fig2}
\end{figure}

To constrain the warm inflation parameter $Q$, we resort to Eq.(\ref{Eq t-to-s ratio}) to illustrate the behavior of respective trajectories in $r-n_s$ plane and compare with Planck observations. Assuming $\lambda=4.78*10^{-79}$ \cite{Garcia}, $T=10^{32}$, $\dot\phi_0=10^{-3}$ and setting different values for $\beta$ and $Q$, we are able to evaluate $r$ in terms of the number of $e$-folds $N$ and plot useful trajectories within $r-n_s$ plane. Our results have been presented in FIG.\ref{fig1} and FIG.\ref{fig2} in which there are two contours that have been obtained from Planck 2018 results \cite{Banerjee}. The smaller and the bigger contours are respectively assigned to 1$\sigma$ and 2$\sigma$ confidence regions.

In FIG.\ref{fig1}, we have illustrated three lines for $Q=1000$, but different values of $\beta$. The black solid line, the red dashed line and the blue dotted line have been plotted for $\beta=0.0074$, $\beta=0.0083$ and $\beta=0.0092$, respectively. In all of them, the upper circle relates to $N=50$ while the lower one relates to $N=70$. It is clear that for $\beta=0.0083$, the trajectory enters the 1$\sigma$ confidence region at $N\approx50$ while this happens for the two others at $N\approx70$. Although all the trajectories lie within the contours thoroughly but the middle case for $\beta=0.0083$, is in a better agreement with the result in \cite{Akrami}. So, we focus on this situation to study the role of $Q$ in our model in the following.

%\begin{table}
%\vspace{.5cm}\hspace{0cm}\label{Table 1}
%\caption[]{The spectral index and tensor to scaler ratio determined from Eqs.(\ref{Eq spectral index}) and (\ref{Eq t-to-s ratio}) for $N=50$, $\beta=0.007$ and $Q=280$.}
%\begin{tabular}{|c||c|c|} \hline
%$\lambda$                               &$r$                 &$n_s$                \\ \hline
%$0.001$                             &$2\times10^{-5}$       &$0.959$            \\
%$10^{-4}$                           &$2\times10^{-6}$       &$0.959$                \\\hline
%\end{tabular}
%\end{table}
%
%\begin{table}
%\vspace{.5cm}\hspace{0cm}\label{Table 2}
%\caption[]{The spectral index and tensor to scaler ratio determined from Eqs.(\ref{Eq spectral index}) and (\ref{Eq t-to-s ratio}) for $N=50$, $\beta=0.007$ and $Q=150$.}
%\begin{tabular}{|c||c|c|} \hline
%$\lambda$                               &$r$                 &$n_s$                \\ \hline
%$0.001$                             &$3\times10^{-4}$       &$0.959$            \\
%$10^{-4}$                           &$3\times10^{-5}$       &$0.959$                \\\hline
%\end{tabular}
%\end{table}

FIG.\ref{fig2} demonstrates three trajectories in the interval $50<N<70$, for $\beta=0.0083$ and different values of $Q$. The black solid line, the red dashed line and the blue dotted line have been plotted for $Q=2400$, $Q=1000$ and $Q=440$, respectively. Obviously we can conclude that for a given value of $\beta$, the greater $Q$ is in more agreement with observations. For example, setting $\beta=0.0083$, for $Q>1000$, the trajectories place within the 1$\sigma$ confidence region for $50<N<70$.

\begin{figure}[h]
\centering
\includegraphics[width=8cm]{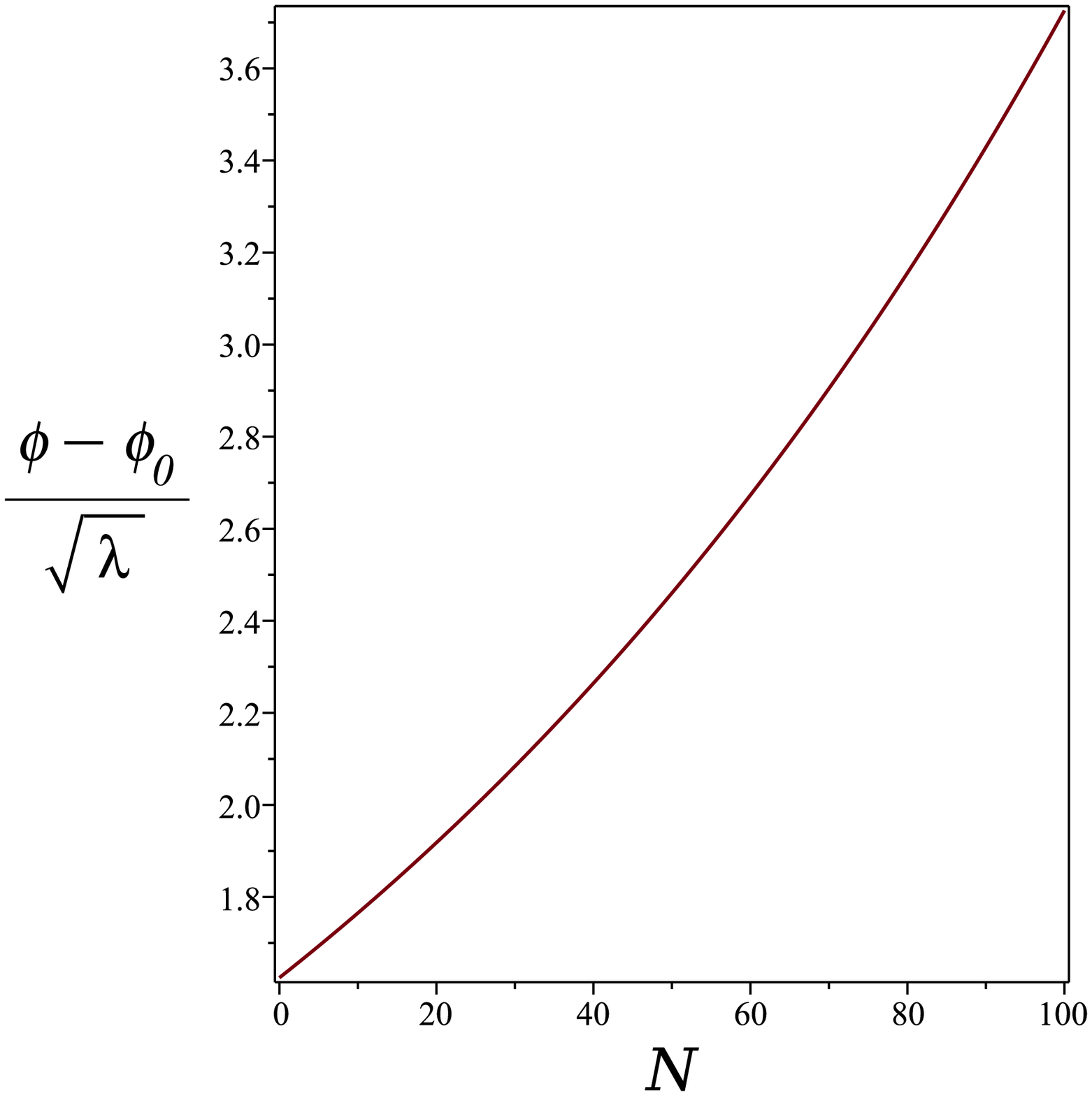}
\includegraphics[width=8cm]{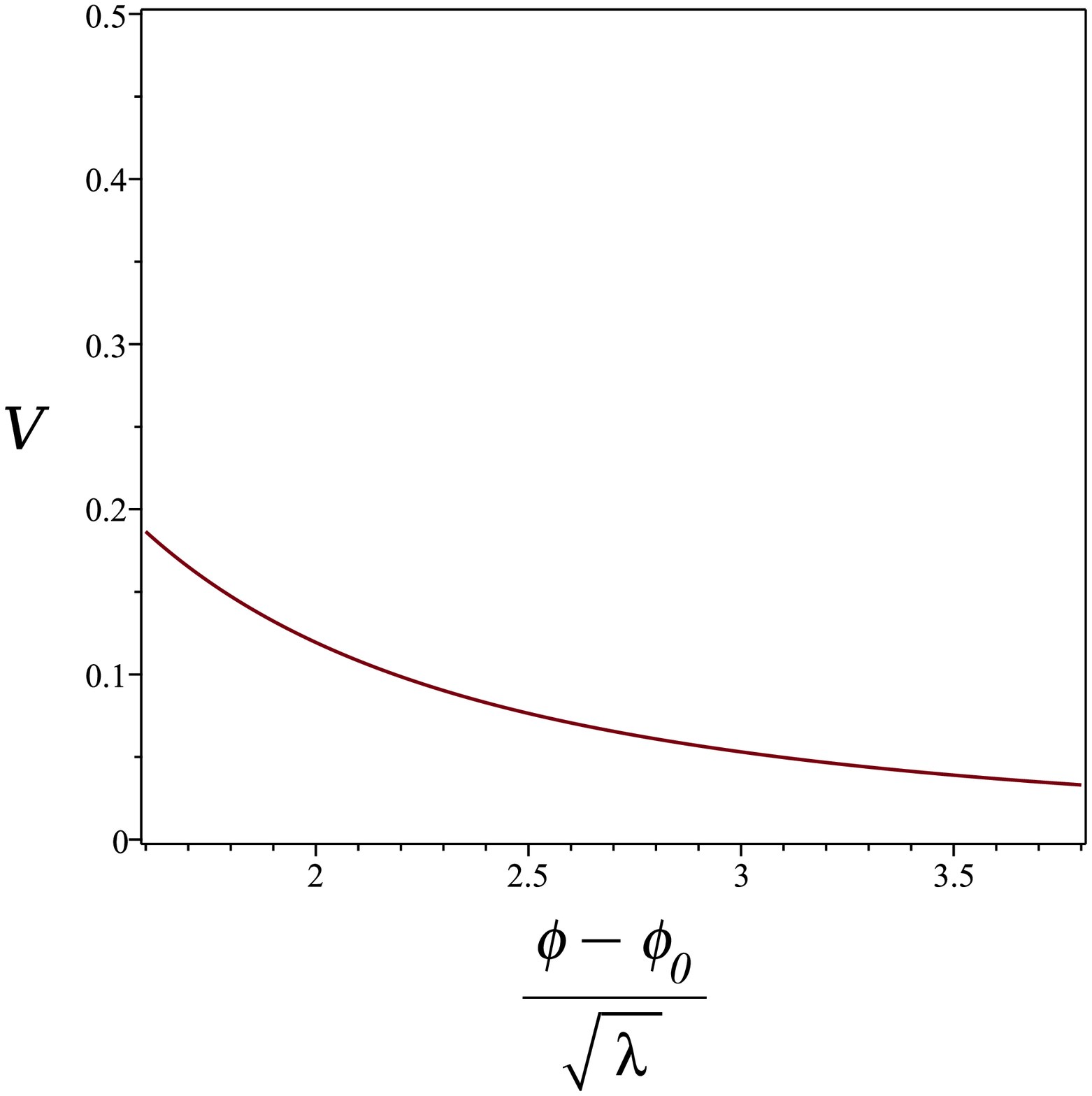}
\caption{Left: the behavior of $(\phi-\phi_0)/\sqrt{\lambda}$ with respect to $N$. The inflaton field grows during inflation. Right: the behavior of $V$ with respect to $(\phi-\phi_0)/\sqrt{\lambda}$. The potential curve is nearly flat that is compatible with the constant-roll inflationary scenario.}\label{fig3}
\end{figure}

FIG.\ref{fig3} illustrates the behavior of the inflaton field with respect to the number of $e$-folds $N$ and also the behavior of its potential in terms of the inflaton field in which we have used $\beta=0.0083$, $Q=1000$ and $\dot\phi_0=10^{-3}$. In the left plot, variation of $(\phi-\phi_0)/\sqrt{\lambda}$ with respect to $N$ has been demonstrated for the interval $0<N<100$. We find that the scalar field increases during inflation, generally. Also, in the right plot the behavior of $V$ with respect to $(\phi-\phi_0)/\sqrt{\lambda}$ has been displayed for the same interval as the left one. This trajectory shows a nearly flat potential that indicates our model approximates a standard slow-roll inflation for small values of $\beta$.

Although to find analytical solutions we only investigated the case $V_0=0$, one can check the situation for the general case $V_0\neq0$, numerically. It is out of the scope of this article but we would like to note that for $V_0\neq0$, along with increasing the value of each of the parameters $V_0$, $\beta$ and $Q$, the inflaton field still increases but much slower than it evolved in the case $V_0=0$. Also, the inflaton potential $V$ gets more flattened in the same interval so that $\partial V/\partial\phi\approx0$. The state that is usually called the ultra-slow-roll inflation.

\section{Conclusions}

In this manuscript we have investigated the warm inflationary Universe under the constant-roll conditions on the brane. In section 2, we have used a novel approach to solve the Friedman equation on the brane considering the constant-roll condition given by $\ddot{\phi}+\beta H\dot{\phi}=0$. The number of $e$-folds, $N$, has been used as the independent variable and the inflaton field, its potential energy and the Hubble parameter has been determined in terms of $N$.

In section 3, the evolution of primordial perturbations has been studied for both scalar and tensor perturbations. We have ignored the case $V_0\neq0$, because the analytical approach is only applicable for $V_0=0$. We have calculated some of the most important perturbation parameters such as the scalar spectral index $n_s$ and the tensor-to-scalar ratio $r$, in terms of model parameters $\beta$ and $Q$.

In section 4, we have constrained our model parameter $\beta$ using the latest observational Planck data for $n_s$. Also, we have illustrated several trajectories in $r-n_s$ plane to constrain the other model parameter $Q$. We have found that our model can be in good agreement with observational data. Irrespective of perturbation parameters we have displayed the behavior of the inflaton field and its potential with respect to $N$, that the latter shows a behavior similar to one of slow-roll inflationary model.

\section{Acknowledgments}

We thank the anonymous referee for important comments.

\section{Appendix}

The following equation is known as the generalized Bessel equation
\begin{equation}\label{Eq generalized Bessel equation}
x^2\frac{d^2y}{dx^2}+(1-2\nu)x\frac{dy}{dx}+(\theta^2\gamma^2x^{2\gamma}+(\nu^2-n^2\gamma^2))y=0,
\end{equation}
and its general solution is given by:
\begin{equation}\label{Eq generalized Bessel equation2}
y(x)=Ax^{\nu}J_n(\theta x^{\gamma})+Bx^{\nu}Y_n(\theta x^{\gamma}).
\end{equation}

\end{document}